\documentclass[aps,pre,reprint,showpacs,superscriptaddress,longbibliography]{revtex4-1}
\usepackage{amsmath,amssymb,graphicx,units}
\usepackage[usenames,dvipsnames]{color}

\begin{document}
\title{Eigenstate Thermalization and Spontaneous Symmetry Breaking
in the One-Dimensional Transverse-Field Ising Model with Power-Law Interactions}
\author{Keith R. Fratus}
\affiliation{Department of Physics, University of California, Santa Barbara, California, 93106, USA}
\author{Mark Srednicki}
\affiliation{Department of Physics, University of California, Santa Barbara, California, 93106, USA}

\begin{abstract}
We study eigenstate thermalization and related signatures of quantum chaos 
in the one-dimensional ferromagnetic transverse-field Ising model with power-law interactions. The presence of long-range interactions allows for a finite-temperature phase transition despite the one-dimensional geometry of the model. Unlike previous studies of eigenstate thermalization in non-disordered systems with finite temperature phase transitions, our model possesses sufficiently many energy eigenstates below the critical energy density to allow us to make a definitive statement about the presence of eigenstate thermalization and chaotic level statistics in the broken-symmetry phase. 
\end{abstract}

\maketitle

\section{Introduction}

The eigenstate thermalization hypothesis (ETH) \cite{deutsch1991quantum, srednicki1994chaos, srednicki95, srednicki99, rigol_dunjko_08} has recently been the subject of a large body of experimental and theoretical work \cite{rigol_09a, rigol_09b, santos_rigol_10b,santos_rigol_10a, neuenhahn_marquardt_12, genway_ho_12, khatami_pupillo_13, beugeling_moessner_14, kim_14, sorg_vidmar_14, AETRigol,FSSIkeda,PLETHSteinigweg,Khlebnikov2013yia,Garrison2015lva,NDPolkovnikov}. ETH can explain how an isolated, quantum many-body system in an initial pure state can come to thermal equilibrium (as determined by measurements of a specified set of observables) in finite time, and is thus fundamental to understanding the validity of conventional quantum statistical mechanics as an accurate description of the long-time behavior of quantum systems; for a review, see \cite{mrigolreview2015}. ETH is expected to hold in systems without disorder that are sufficiently far from integrability 
(including effective integrability caused by many-body localization in disordered systems), 
for observables that are sufficiently simple (e.g. local) functions of the fundamental degrees of freedom.

The key statement of ETH is that expectation values of a relevant
observable $M$ in an energy eigenstate $|\alpha\rangle$
(of the full many-body hamiltonian $H$) take the form
\begin{equation}
\langle\alpha|M|\alpha\rangle = {\cal M}(E_\alpha),
\label{M}
\end{equation}
where ${\cal M}(E)$ is a smooth function of $E$ and $E_\alpha$ is the energy eigenvalue. 
In a system with $N\gg 1$ degrees of freedom,
this is enough information to show that 
${\cal M}(E)$ is equal, up to $O(N^{-1/2})$ corrections, to the canonical
thermal average of the operator $M$,
\begin{equation}
{\cal M}(E) = {\mathop{\rm Tr} M e^{-H/kT}\over \mathop{\rm Tr} e^{-H/kT}}
\bigl[1 + O(N^{-1/2})\bigr],
\label{ME}
\end{equation}
where the temperature $T$ is implicitly determined as a function of energy $E$ by
the usual relation
\begin{equation}
E = {\mathop{\rm Tr} H e^{-H/kT}\over \mathop{\rm Tr} e^{-H/kT}}.
\label{E}
\end{equation}

A second key statement of ETH is that the off-diagonal matrix elements of $M$ in
the energy basis, $\langle\alpha|M|\beta\rangle$ with $\alpha\ne\beta$,
are exponentially small in $N$.  This is needed to explain why the diagonal
matrix elements of Eq.~(\ref{M}) dominate the instantaneous expectation value
of $M$ (in a generic time-dependent state) at almost all times, 
which in turn is necessary for thermal equilibrium to be maintained
once it has been achieved. However this aspect of ETH will not be our focus.

Eigenstate thermalization is also closely related to the subject of quantum chaos, and many previous numerical studies have found that the onset of quantum chaos, as diagnosed by the 
level-spacing statistics for the energy eigenvalues matching those of the Gaussian orthogonal
ensemble (GOE) of random matrices, is typically associated with the onset of eigenstate thermalization \cite{mrigolreview2015,santos_rigol_10b,santos_rigol_10a}.

Some recent work \cite{HuseSpecies,FratusSrednicki,ETHMFSR} has focused on the compatibility of ETH with another paradigm of condensed matter physics, spontaneous symmetry breaking (SSB) and long-range order. For both clean and disordered systems, compatibility between ETH and SSB has been observed in these studies. However, one of the limitations of previous studies of ETH in clean systems with SSB has been the inability to robustly verify 
the predictions of eigenstate thermalization and quantum chaos strictly within the broken symmetry phase, largely due to the relatively small number of energy eigenstates in this 
regime for the finite-size systems that are amenable to exact diagonalization.
Previous work for clean systems has been on the transverse field Ising model (TFIM) in two space dimensions,
but the largest tractable lattice size has been $4\times5$ \cite{FratusSrednicki,ETHMFSR}, which turns out to have only a small number of states (as few as one or two, depending on the strength of the transverse field) in the broken-symmetry phase.

Any exact-diagonalization study of a quantum Ising system is limited by the total number of spins. Arranging the spins in a one-dimensional lattice results in the largest possible linear dimension for a model with a fixed number of spins, and thus provides the best possible geometry for attempting to resolve the details of a finite temperature phase transition. 
We therefore seek a computationally tractable Ising model in one space dimension.
As a result of the Mermin-Wagner theorem, however, any finite temperature phase transition in a one-dimensional model with local interactions is forbidden, but this is not the case
for interactions that fall off as a power of the distance between spins \cite{Dyson1969}. Therefore, we study the one-dimensional, ferromagnetic transverse field Ising model with power-law interactions,
a quantum many-body system that possesses a finite-temperature phase transition \cite{PTDutta,FRSGlumac,Sergio97,Kosterlitz76,Barati2000,Cannas95,Chandra2010,Vodola16,Sandvik2003}. While previous studies \cite{antonio2012,Halimeh2016,Li2016} have investigated the subject of thermalization in systems with long-range interactions, we focus specifically on the question of finite-temperature, spontaneous symmetry breaking, and its compatibility with ETH.
Our results confirm the existence of eigenstate thermalization in this model, as well as chaotic level statistics, within the broken-symmetry phase.

This paper is organized as follows. 
In Sec.~\ref{sec:sec2}, we present the details of the specific model we study and
the numerical approach that we use. 
In Sec.~\ref{sec:sec3}, we give our numerical results for this model.
In Sec.~\ref{sec:sec4}, we discuss the evidence for ETH and quantum chaos.
In Sec.~\ref{sec:sec5}, we briefly discuss the implications that our results have for time-evolution in this model.
We conclude in Sec.~\ref{sec:sec6}.

\section{Model and Numerical Approach}\label{sec:sec2}

Our model Hamiltonian can be written as
\begin{equation}
 \hat H = -\sum_{i \neq j}J_{ij}\hat\sigma_{ i}^z\hat\sigma_{ j}^z - g\sum_{ i}\hat\sigma_{ i}^x ,
 \label{eq:hamiltonian}
\end{equation}
where $\sigma_{ i}^z$ and $\sigma_{ i}^x$ are the standard Pauli matrices on site $i$ of the one-dimensional lattice. The Ising interaction $J_{ij}$ is chosen to obey a power-law decay,\begin{equation}
 J_{ij} = \frac{J}{|i-j|^{p}}.
\end{equation}
We set $J=1$, which fixes the energy scale, and corresponds to a ferromagnetic Ising coupling. 
For the transverse term, we choose $g = 1.5$. This value is roughly half-way between the integrable limit at $g=0$, where we do not expect to see any eigenstate thermalization or quantum chaos, and the point at which there is a quantum phase transition, where there is no longer any order at any temperature. A combination of Quantum Monte Carlo and Mean-Field calculations lead us to believe that this quantum phase transition is somewhere between $g=3.5$ and $g=4.0$ (a knowledge of the precise location of this quantum phase transition is not necessary for our purposes). Our boundary conditions are chosen to be open, so we do not make use of translation symmetry in diagonalizing the Hamiltonian. We do, however, make explicit use of spatial parity symmetry and Ising symmetry. Unlike previous studies \cite{FratusSrednicki}, we do not include any explicit symmetry breaking term, and instead focus on observables which are invariant under the $Z_2$ Ising symmetry. 

For values of the exponent $p < 1$, the long-range interactions between spins are powerful enough to destroy extensivity, thus precluding the existence of a well-defined thermodynamic limit. For $p > 2$, the long-range interactions are weak enough such that there is no finite-temperature phase transition. For $1 < p < 2$, both a finite-temperature phase transition and a well-defined thermodynamic limit exist \cite{Dyson1969}, and hence this is the parameter range in which we are interested. In this work we choose $p = 1.5$. 

In our work, the largest system size for which we are able to find exact eigenstates has N = 27 Ising spins. The full Hilbert space of this 27-site model contains 134,217,728 states, while the even-parity, even-Ising mode contains 33,558,528 states. Since a Hilbert space of 33 million states is much too large to fully diagonalize with current technology, we instead find only the 250 lowest-energy states, using a standard Lanczos treatment. Since we are interested only in studying the behavior of eigenstates in the low-energy, broken-symmetry phase, this is sufficient for our purposes. To compare our exact diagonalization data with predictions from a standard canonical ensemble, we also perform a Stochastic Series Expansion (SSE) Quantum Monte Carlo (QMC) calculation, using a technique similar to the one in Ref. \cite{Sandvik2003}. This allows us to compare against the standard thermal prediction for both the 27-site system (which we cannot compute exactly since we lack information on the full spectrum), and also for much larger system sizes, which provides insight into the behavior of this system in the thermodynamic limit.

\section{The Broken-Symmetry Phase}\label{sec:sec3}

We begin by motivating the claim that we are able to study eigenstates which lie within the broken-symmetry (ordered) phase.
We examine the behavior of three quantities: the ferromagnetic structure factor, the Binder cumulant, and the full
probability distribution of the magnetic order parameter.

We begin by studying the behavior of the Binder cumulant, which is defined as
\begin{equation}
U \equiv 1 - \frac{\langle M_{z}^{4} \rangle}{3 \langle M_{z}^{2} \rangle^{2}},
 \label{eq:binder}
\end{equation}
where
\begin{equation}
\hat M_{z} \equiv \sum_{ i}\hat\sigma_{ i}^z
 \label{eq:mz}
\end{equation}
is the order parameter, and the angle brackets refer either to averaging with respect to the canonical ensemble, or the expectation value within an energy eigenstate, which are the same in the case that ETH is satisfied. The Binder cumulant quantifies the extent to which the full probability distribution of the order parameter $M_{z}$ reflects the behavior of the 
ordered or non-ordered
phases \cite{Binder81}. At low temperatures, the full probability distribution approaches two well-separated Gaussian distributions at equal and opposite non-zero values of the magnetization. In this limit, the Binder cumulant approaches a value of $2/3$, up to corrections which scale as $1/N$. At high temperatures, the full probability distribution approaches a single Gaussian distribution 
peaked around zero net magnetization. In this limit, the Binder cumulant approaches a value of zero, again
up to corrections which scale as $1/N$. In the large system size limit, the transition between these two Binder cumulant values is sharp, with a value at the critical temperature, $U \left ( T_{c} \right )$.
When the Binder cumulant is plotted as a function of temperature, the crossing point for different system sizes provides a good estimate for the critical temperature. 

\begin{figure}[h]
\centering
\includegraphics[width=85mm]{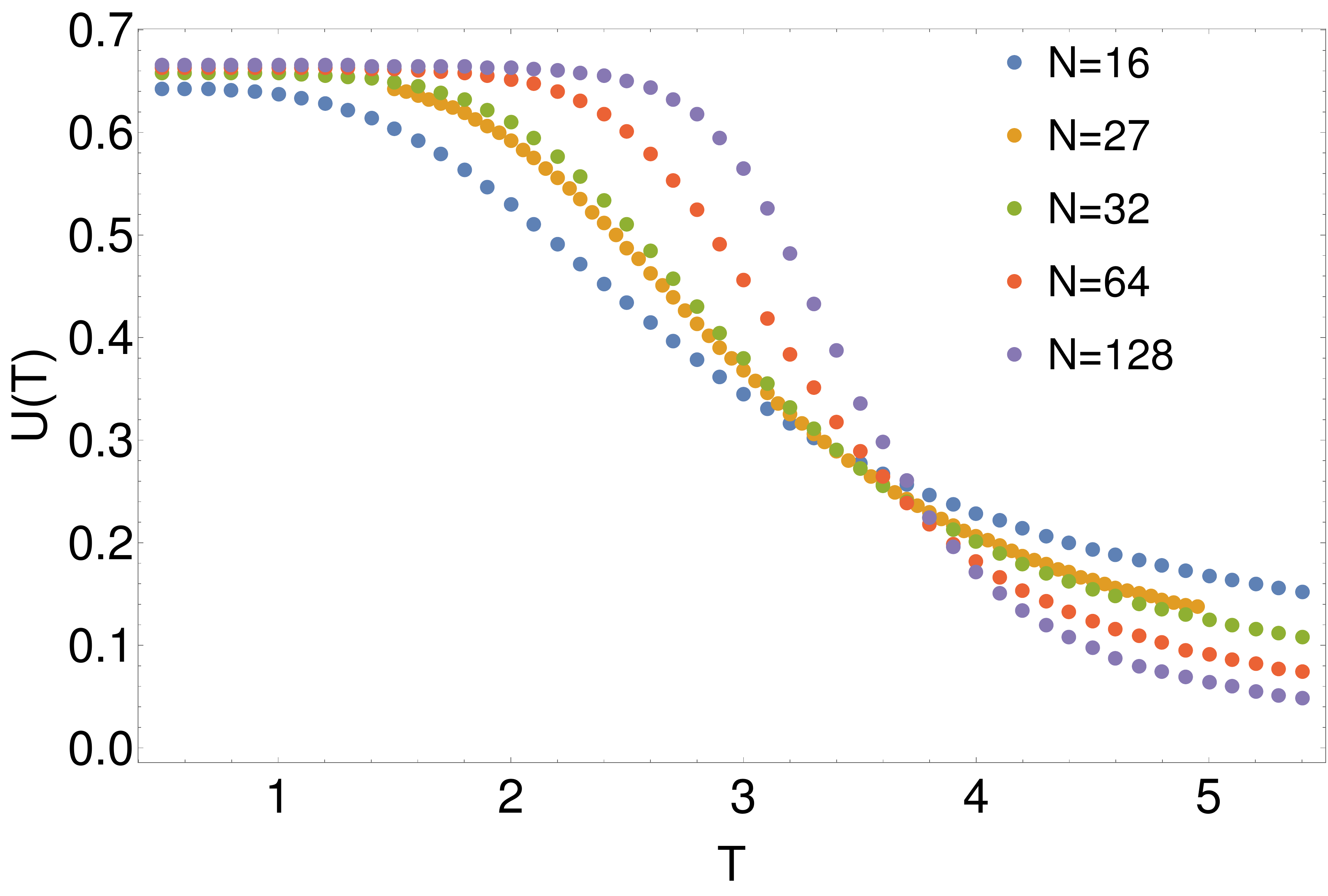}
\caption{The Binder cumulant as a function of temperature, for system sizes N = 16 (blue), 27 (yellow), 32 (green), 64 (orange), and 128 (purple).}
\label{fig:bctemp}
\end{figure}

To provide context for the results we find from exact diagonalization, Figure \ref{fig:bctemp} shows a plot of the Binder cumulant as a function of temperature in the canonical ensemble, for various system sizes, computed using SSE. By  
examining the crossing point of the 27 and 32 site models, we find 
$T_{c} \approx$ 3.53. At this temperature, the energy density of the 27-site model is 
$E_{c}/N \approx$ -1.08. This energy density is well above the range 
that we will consider when 
we construct the low-lying energy eigenstates by exact diagonalization.

\begin{figure}[h]
\centering
\includegraphics[width=85mm]{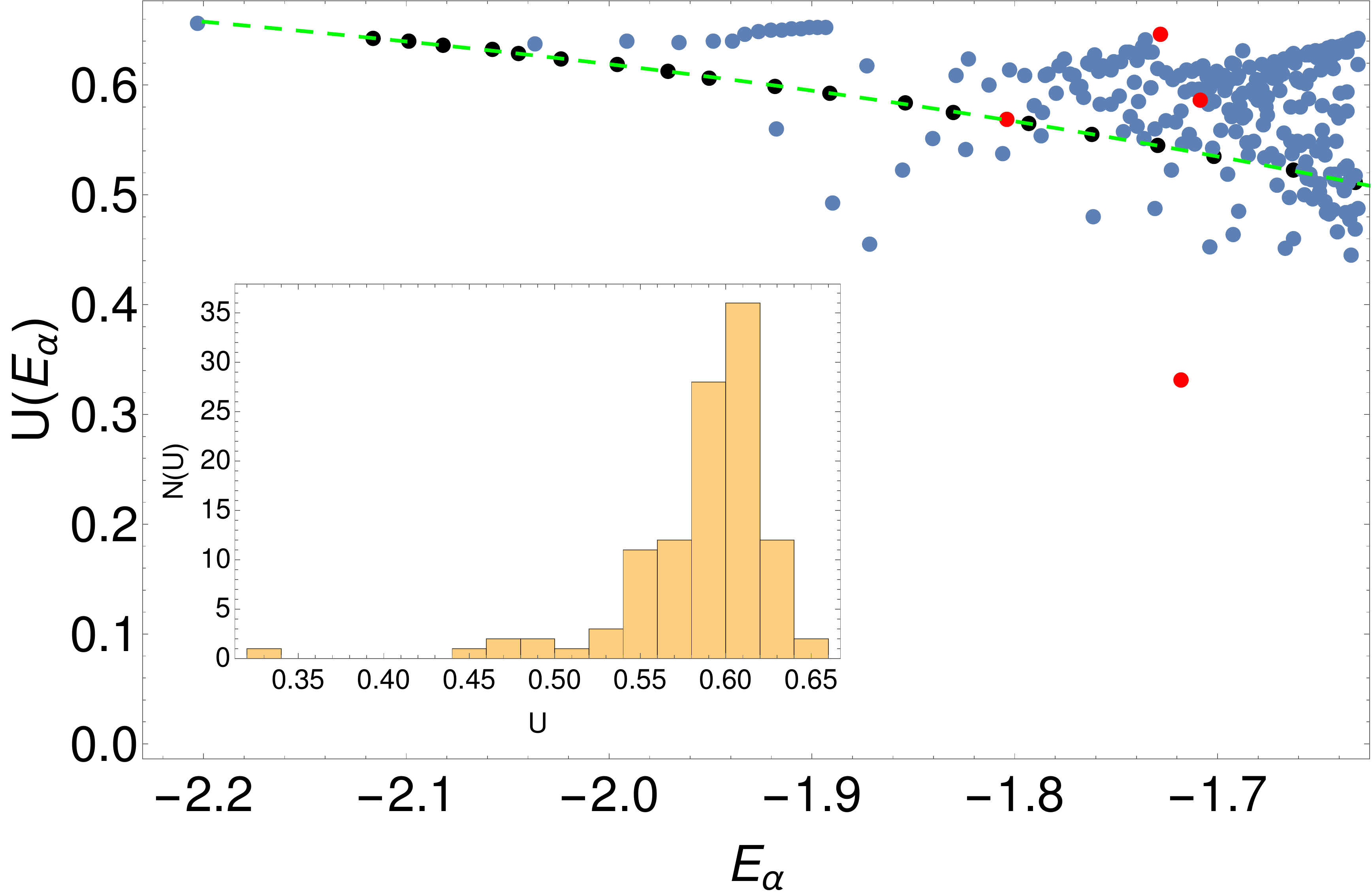}
\caption{The Binder cumulant as a function of energy for the 27-site system, with exact diagonalization data in blue, and SSE data in black. The points highlighted in red correspond to the states which are studied in more detail in Figure \ref{fig:4panelpdf}. The green dashed line is a quartic polynomial fit to the SSE data. Inset is a Histogram of the Binder cumulant values between an energy density of -1.82 and -1.68, the energy density in which we extract the level spacing statistics.}
\label{fig:bc}
\end{figure}

For the exact-diagonalization data, extracted using the lowest 250 eigenstates of the 27-site model, a plot of the Binder cumulant as a function of energy is shown in figure \ref{fig:bc};
$\langle M_{z}^{2} \rangle$ and $\langle M_{z}^{4} \rangle$ are 
eigenstate expectation values. Overlayed on the plot is the Binder cumulant as a function of energy density, as computed using SSE.  A histogram of Binder cumulant values, between an energy density of $-1.82$ and $-1.68$, is inset. This is the energy density range for which we will later extract the level spacing statistics of this model. The Binder cumulant values are distributed close to, but not exactly around, a value of $2/3$. Specifically, the mean value of the Binder cumulant over this energy density window is given by $\overline{U} = 0.586$, with a standard deviation of $0.044$. 

In addition to displaying the Binder cumulant, we also examine the full probability distribution (FPD) of $M_{z}$ in exact energy eigenstates. It can be shown \cite{srednicki99} that if an observable satisfies ETH, any multiplicative power of that observable must also necessarily satisfy ETH. Since any probability distribution with well-defined moments can be reconstructed from these moments, the satisfaction of ETH for all powers of an observable implies that the exact eigenstate FPD of any observable which satisfies ETH must necessarily agree with the thermal prediction. 

\begin{figure}[h]
\centering
\includegraphics[width=42mm]{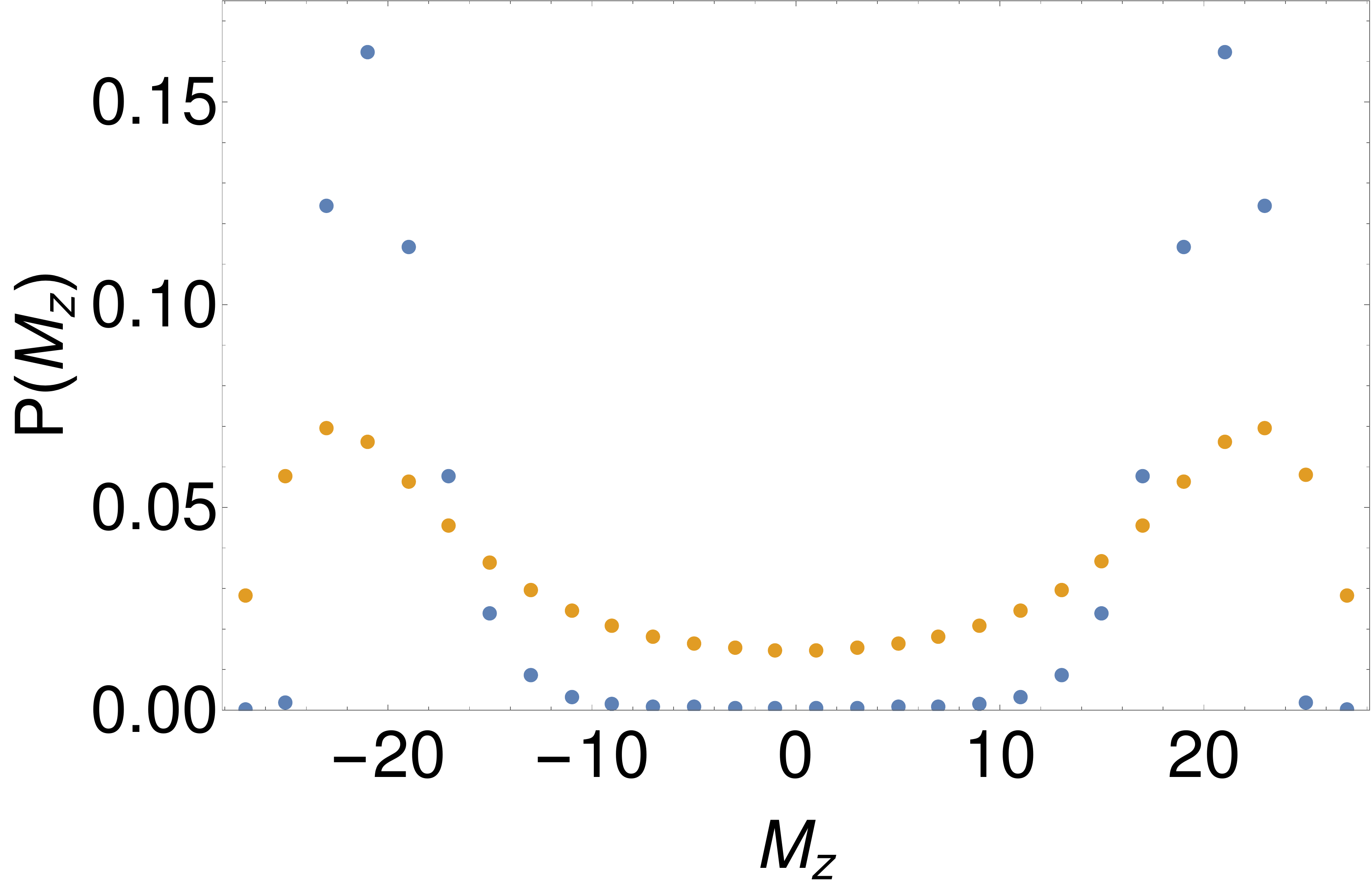}
\includegraphics[width=42mm]{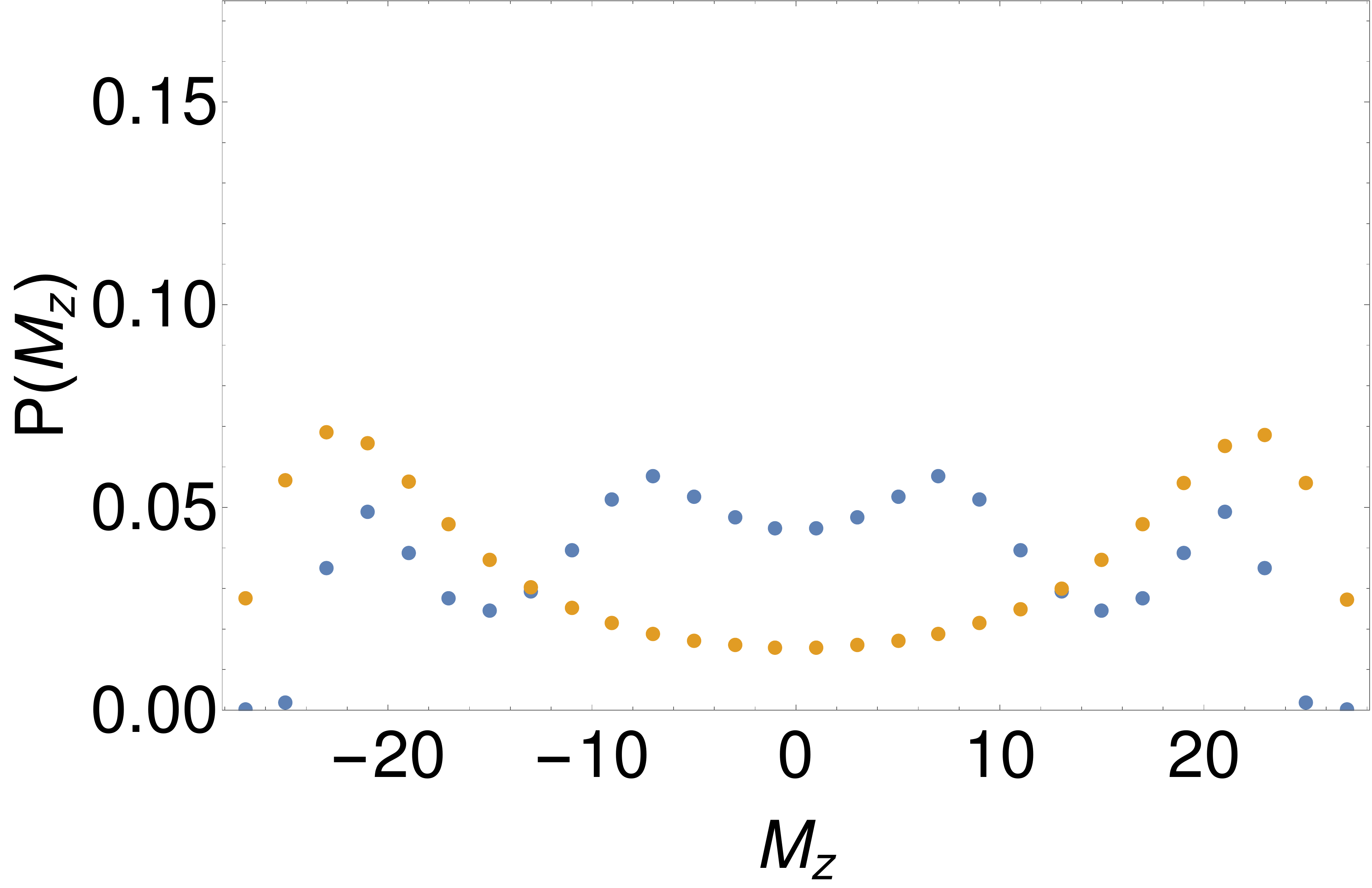} \\
\includegraphics[width=42mm]{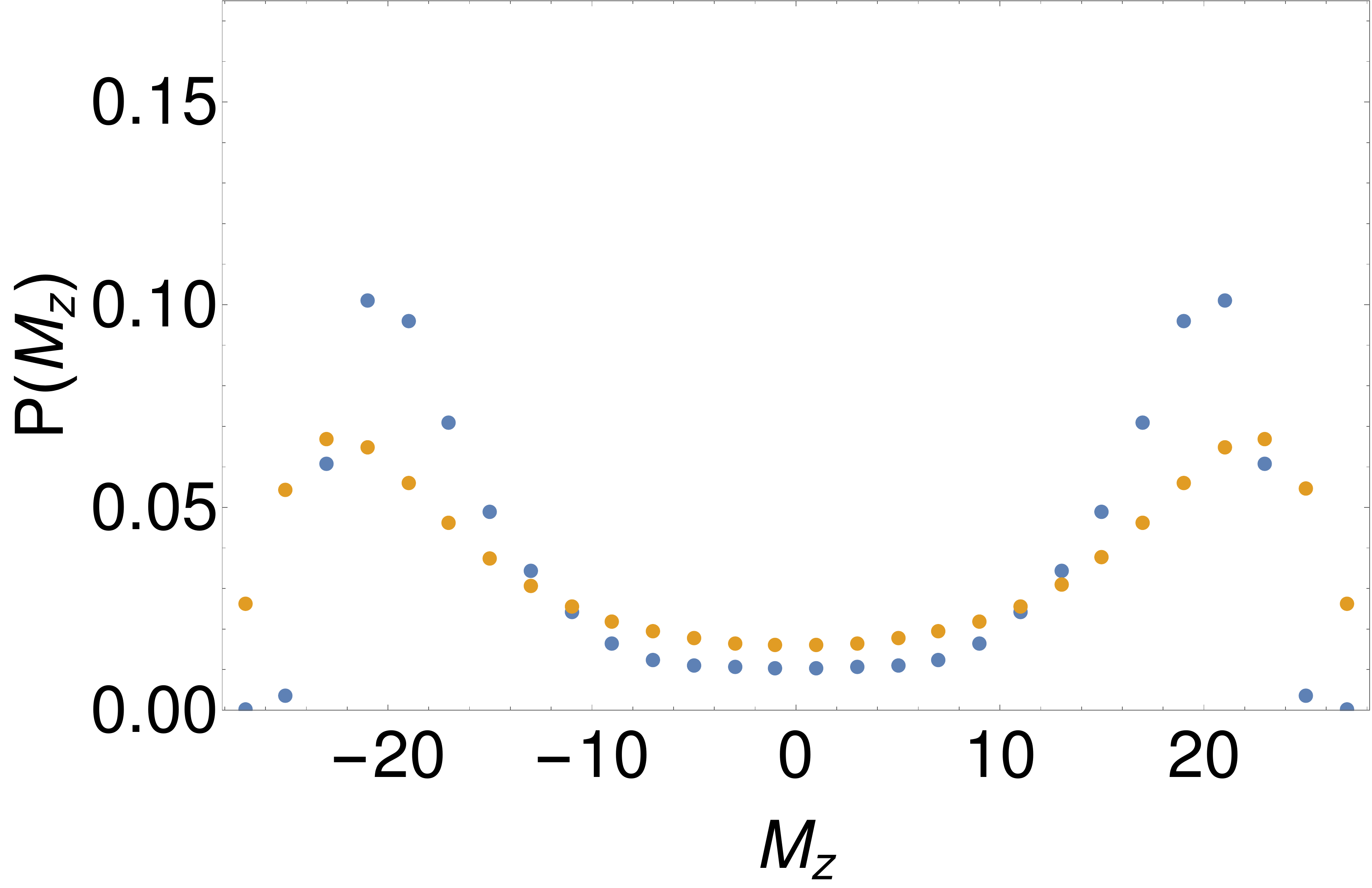}
\includegraphics[width=42mm]{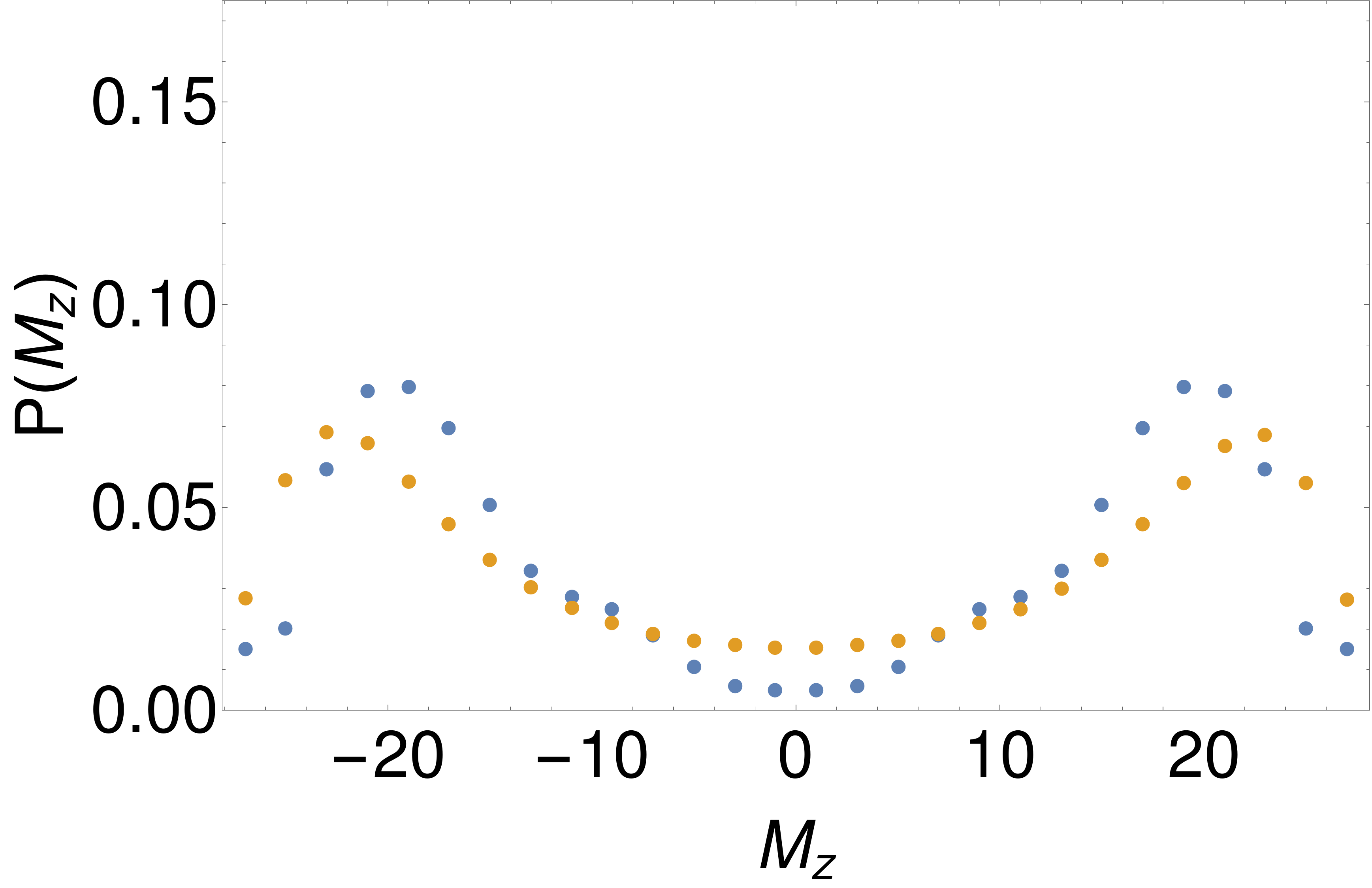}
\caption{The probability distribution for $M_z$, shown in blue, in the eigenstates with energy density $-1.72845$ (top left), $-1.71831$ (top right), $-1.70854$ (bottom left), and $-1.80384$ (bottom right). The orange points indicate the canonical ensemble prediction for the probability distribution at the same energy scale.}
\label{fig:4panelpdf}
\end{figure}

We display the FPD of $M_{z}$ for several representative energy eigenstates in Figure \ref{fig:4panelpdf}. The upper left panel shows the FPD for the state with energy density -1.72845, for which the Binder cumulant attains a value of 0.646, the closest to 2/3 of any eigenstate. The behavior of this FPD clearly resembles that of two well-separated peaks, at equal and opposite magnetizations. However, despite the fact that this eigenstate has a Binder cumulant close to 2/3, the value of the Binder cumulant at this energy density in the canonical ensemble is lower, at approximately 0.544. Correspondingly, the probability distribution for this state is more sharply peaked than the probability distribution we would expect in the canonical ensemble, which is also shown in the Figure. This discrepancy is due to both differences between the microcanonical and the canonical ensemble in a small system, as well as a deviation from perfect eigenstate thermalization.

The upper right panel of Figure \ref{fig:4panelpdf} shows the FPD for the state with energy density $-1.71831$, for which the Binder cumulant attains a value of 0.331, the furthest from 2/3 of any eigenstate in this energy window, and additionally the furthest of any state in this window from the thermal prediction at its corresponding energy density. At this energy density, the thermal prediction for the Binder cumulant would be approximately 0.541. The behavior of this FPD is a poor reflection of two well-separated peaks, which is not surprising, given the value of the Binder cumulant. The FPD also deviates noticeably from that predicted by the canonical ensemble. This state is a rare exception, as can be seen in the histogram of Binder cumulant values.

The bottom left panel of Figure \ref{fig:4panelpdf} shows the FPD for the state with energy density $-1.70854$, for which the Binder cumulant attains a value of 0.586, the closest to its average across this energy window. The FPD displays two well-separated peaks, however there is a non-zero plateau near $M_z = 0$. While the Binder cumulant for this state is a good approximation to the average value across the relevant energy window, it is not necessarily in agreement with the thermal prediction at that energy density, which is approximately 0.537. For this reason, the probability distribution for this state is somewhat more sharply peaked than the thermal prediction.

Lastly, the bottom right panel of Figure \ref{fig:4panelpdf} shows the FPD for the state with energy density -1.80384, for which the Binder cumulant attains a value of 0.569, the closest of any eigenstate to the thermal value of the Binder cumulant at the corresponding energy density, 0.567. The FPD displays two well-separated peaks, however there is a non-zero plateau near $M_{z} = 0$ for the thermal prediction, which is absent in the exact energy eigenstate. This indicates that even for energy eigenstates whose Binder cumulant is very close to the thermal prediction, higher order moments of the probability distribution can still differ noticeably. We expect that as a result of eigenstate thermalization, in the thermodynamic limit all moments of the probability distribution would be equal to their thermal values \cite{srednicki99}.

Taken together, these plots indicate that the vast majority of energy eigenstates in this energy range have a magnetization probability distribution which clearly resembles that of two well-separated peaks. The agreement with the canonical ensemble varies, and there exist states which are rare exceptions to this behavior, but most states capture the correct qualitative behavior of the broken-symmetry phase.

Lastly, we study the finite-size scaling behavior of the ferromagnetic structure factor,
\begin{equation}
 \hat S _{F} \equiv \frac{1}{N} \hat M_{z}^{2}.
 \label{eq:struct}
\end{equation}
Taking the expectation value and using Eq.~(\ref{eq:mz}), we have

\begin{equation}
\langle \hat S _{F} \rangle 
= 1 + \frac{1}{N}\sum_{i \neq j}\langle \hat\sigma_{ i}^z\hat\sigma_{ j}^z \rangle.
 \label{eq:structexpand}
\end{equation}
In the symmetric phase, the behavior of the spin-spin correlation obeys
\begin{equation}
\lim_{|i-j| \to \infty}   \langle \hat\sigma_{ i}^z\hat\sigma_{ j}^z \rangle\sim\exp \left ( -|i-j| / \xi  \right )
 \label{eq:correlabove}
\end{equation}
for some correlation length $\xi$, while in the broken symmetry phase, it obeys
\begin{equation}
\lim_{|i-j| \to \infty}   \langle \hat\sigma_{ i}^z\hat\sigma_{ j}^z \rangle
= c
\end{equation}
for a constant $c$ that is independent of $N$, and is related to the thermodynamic expectation
value of $\hat M_z$ in the broken phase by $c = \langle \hat M_z\rangle^2/N^2$. For this reason, the value of $\langle \hat S _{F}\rangle-1$ 
scales independently of $N$ in the symmetric phase, and linearly with $N$ in the broken symmetry phase (due to the extra factor of $N$ acquired by the double summation in this phase). 
Therefore the ferromagnetic structure factor is also a useful diagnostic for identifying the location of the broken symmetry phase.

\begin{figure}[h]
\centering
\includegraphics[width=85mm]{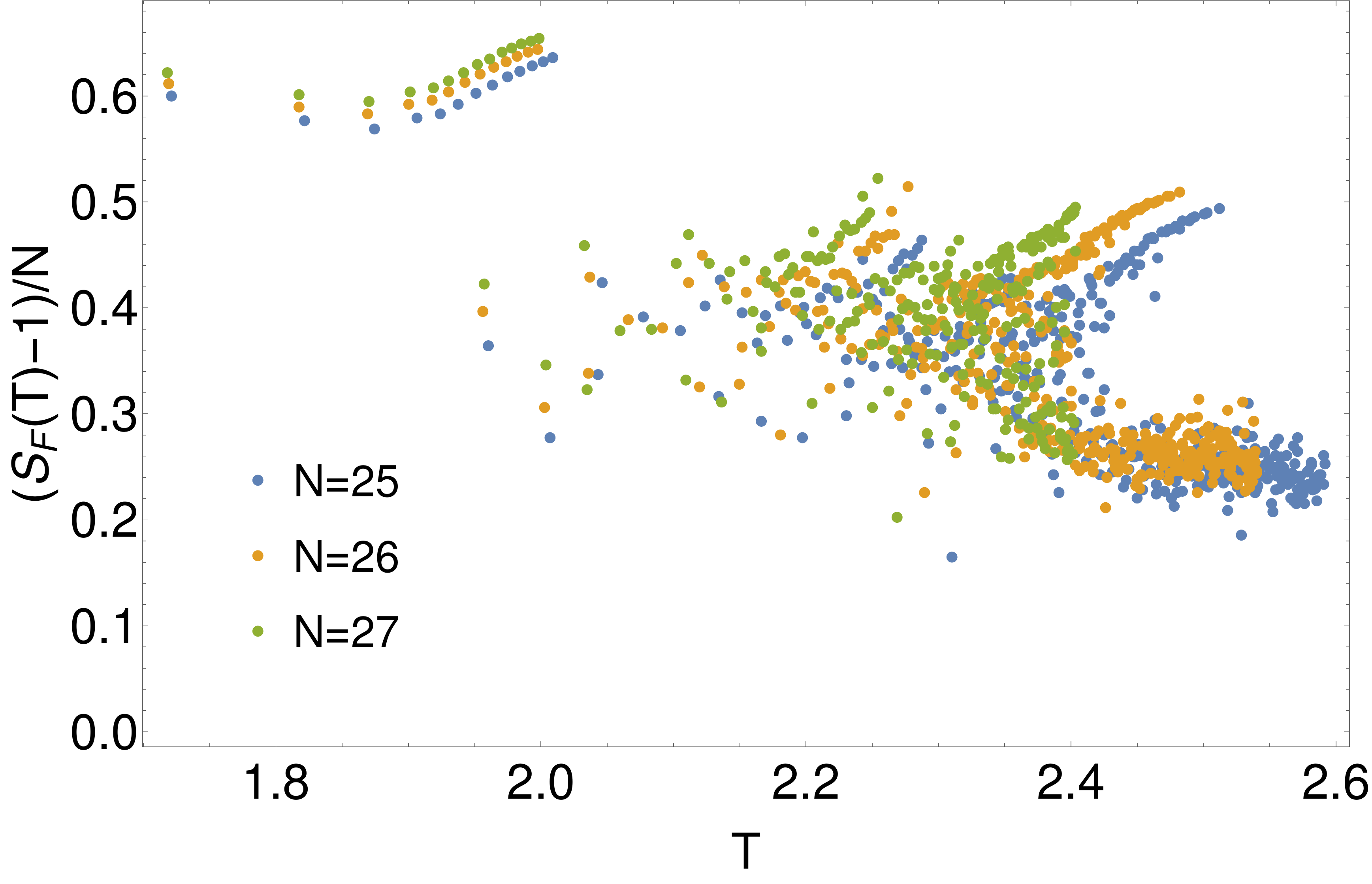}
\caption{The ferromagnetic structure factor for the three largest system sizes we are able to diagonalize, 25 sites (blue), 26 sites (orange), and 27 sites (green), plotted as a function of effective temperature. The ground states are omitted, since they are all trivially found at T=0.}
\label{fig:SFtemp}
\end{figure}

\begin{figure}[h]
\centering
\includegraphics[width=85mm]{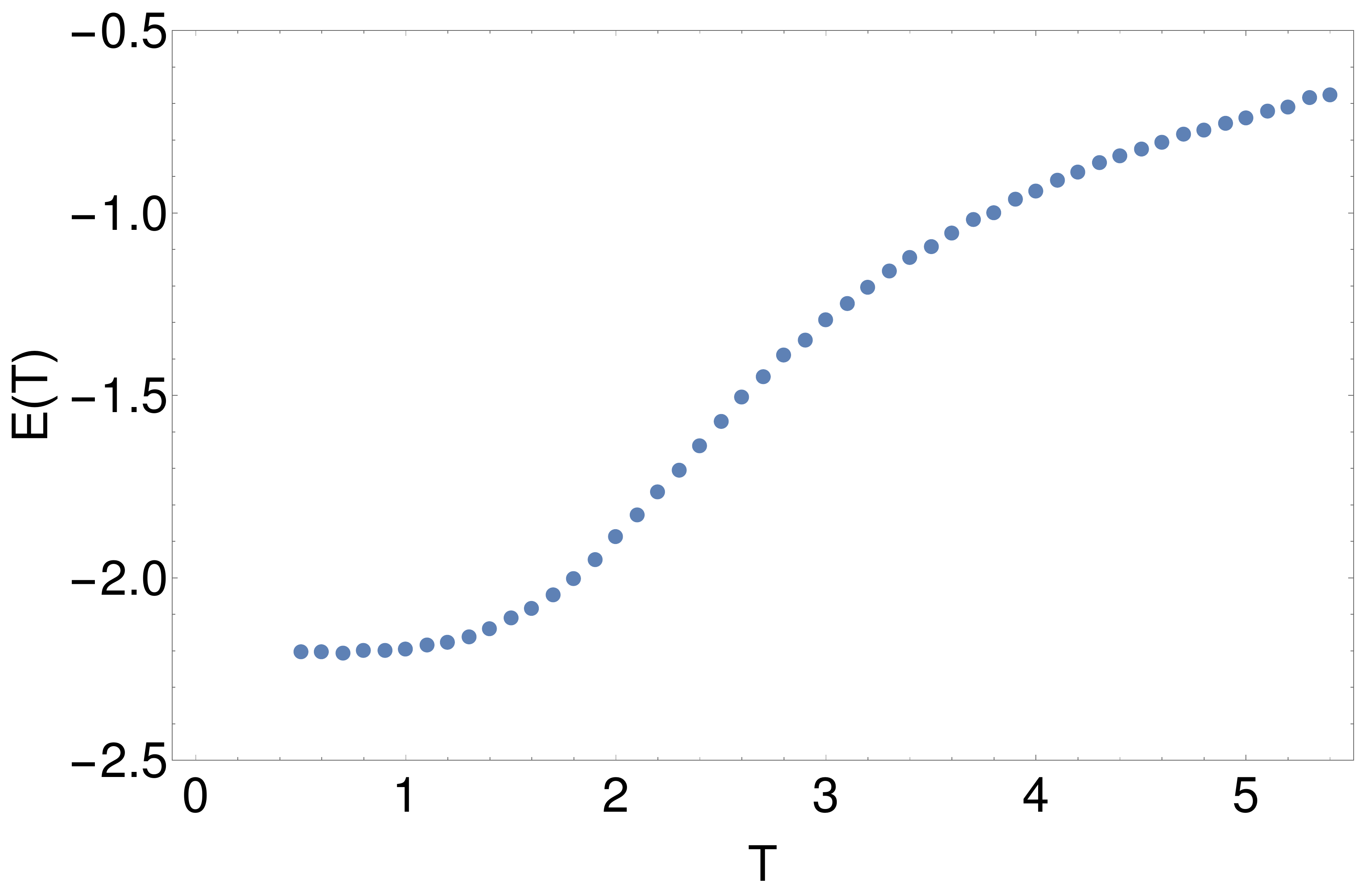}
\caption{A plot of the energy density of the 27-site model as a function of temperature, as computed in the canonical ensemble using the SSE method.}
\label{fig:EvsT}
\end{figure}

In Figure \ref{fig:SFtemp} we display our results for the quantity $(\langle \hat S _{F}\rangle-1)/N$, evaluated in the individual energy eigenstates of our model. While we do not see particularly good satisfaction of ETH by the ferromagnetic structure factor (perhaps demonstrating serious finite size effects for this particular choice of observable), we do note that the data for the structure factor collapses for different system sizes in a way which is still consistent with the symmetry broken phase. This is a marked difference from the results found in our previous work \cite{ETHMFSR}, in which the behavior of the structure factor was consistent with, at most, a handful of energy eigenstates living within the symmetry broken phase. The structure factors in Figure \ref{fig:SFtemp} are plotted as a function of effective temperature, which is computed for a given energy eigenstate by comparing with the temperature in the canonical ensemble which would reproduce the same average energy. We display the data in this manner due to the non-extensive behavior of the Ising term which occurs in this model for small system sizes, which causes the data for different system sizes to align poorly in the horizontal direction when plotted as a function of energy. This mapping between energy and temperature in the canonical ensemble is displayed in Figure \ref{fig:EvsT}.

Taken together, we believe that while our 27 site model still displays noticeable finite-size effects, the behavior of the eigenstates in the energy window we are studying is indicative of the behavior one should expect in the broken symmetry regime.

\section{Quantum Chaos and Eigenstate Thermalization}\label{sec:sec4}

We argue that up to corrections which one should expect for a small system, the behavior of the Binder cumulant as a function of energy, as well as the approximate agreement between the eigenstate probability distributions and the corresponding thermal ones, provide evidence for eigenstate thermalization in the energy range in which we are interested

\begin{figure}[h]
\centering
\includegraphics[width=85mm]{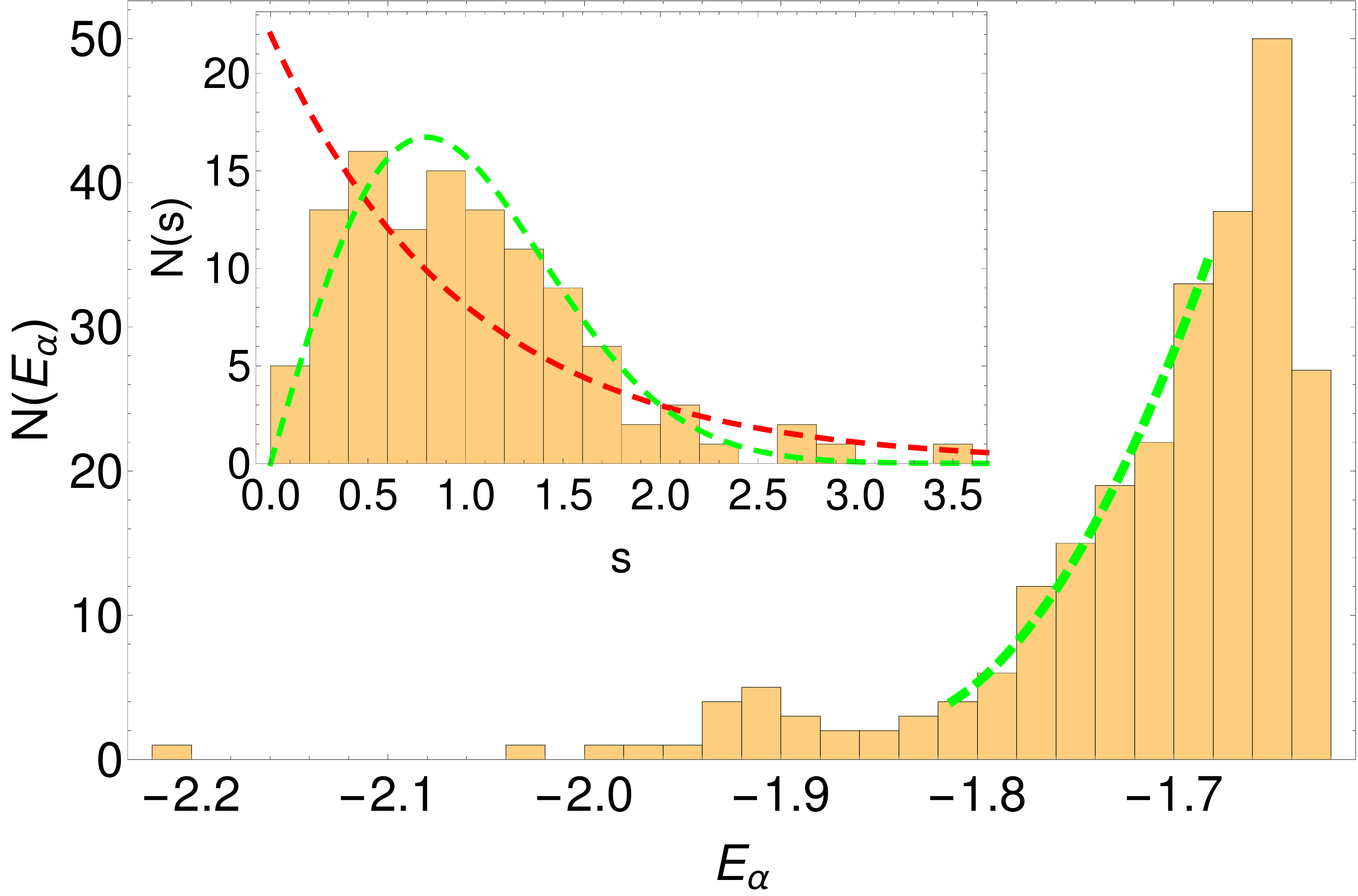}
\caption{The density of states in the 27 site model, as a function of energy density. The green dashed line indicates the curve of best fit used in computing the mean level spacing. The inset shows the level spacing statistics, for energy eigenstates with an energy density between -1.82 and -1.68. The green curve in the inset represents the GOE prediction, while the red curve represents the Poisson prediction.}
\label{fig:LSDOS}
\end{figure}

The main quantum chaos result which we now wish to display is the level spacing statistics of our 27-site model. A histogram of the density of states in our model is shown in Figure \ref{fig:LSDOS}, along with a fit which we use to extract the mean level spacing. In studying the level statistics, we focus on the 111 energy eigenstates which fall between an energy density of -1.82 and -1.68. We choose the lower bound of this window in order to avoid the cluster of low-lying states for which the density of states does not grow exponentially. We choose the upper bound on this window because it is the value for which we have the best agreement with GOE statistics. Interestingly, as we increase the upper bound of the energy density window, despite having a larger sample of level spacings to work with, the agreement between the level spacing statistics and the prediction for the Gaussian ensemble becomes worse. We believe this is a result of the approaching cross-over to the symmetric phase upon approaching higher energy densities. Thus, the upper limit on this window is ultimately chosen to be low enough in energy that we can be confident we have not begun to sample states which reflect the physics of the symmetric phase.

The level spacing statistics themselves are displayed in the inset to Figure \ref{fig:LSDOS}. The dashed green line represents the prediction of the Gaussian Orthogonal Ensemble (GOE) \cite{bohigas_giannoni_84},
\begin{equation}
P(s) = \frac{\pi}{2}s\exp\left(-\frac{\pi}{4} s^2 \right),
\label{eq:goe}
\end{equation}
where $s$ represents the spacing between two energy eigenstates in units of mean level spacing at that energy scale. 
The dashed red line represents the prediction from Poisson level statistics \cite{berry1977level},
\begin{equation}
P(s) = \exp(-s).
\label{eq:poisson}
\end{equation}
The agreement with the GOE prediction is quite good, but as mentioned previously, the agreement has been seen to decrease upon increasing the upper limit of the energy density window.

\begin{figure}[h]
\centering
\includegraphics[width=85mm]{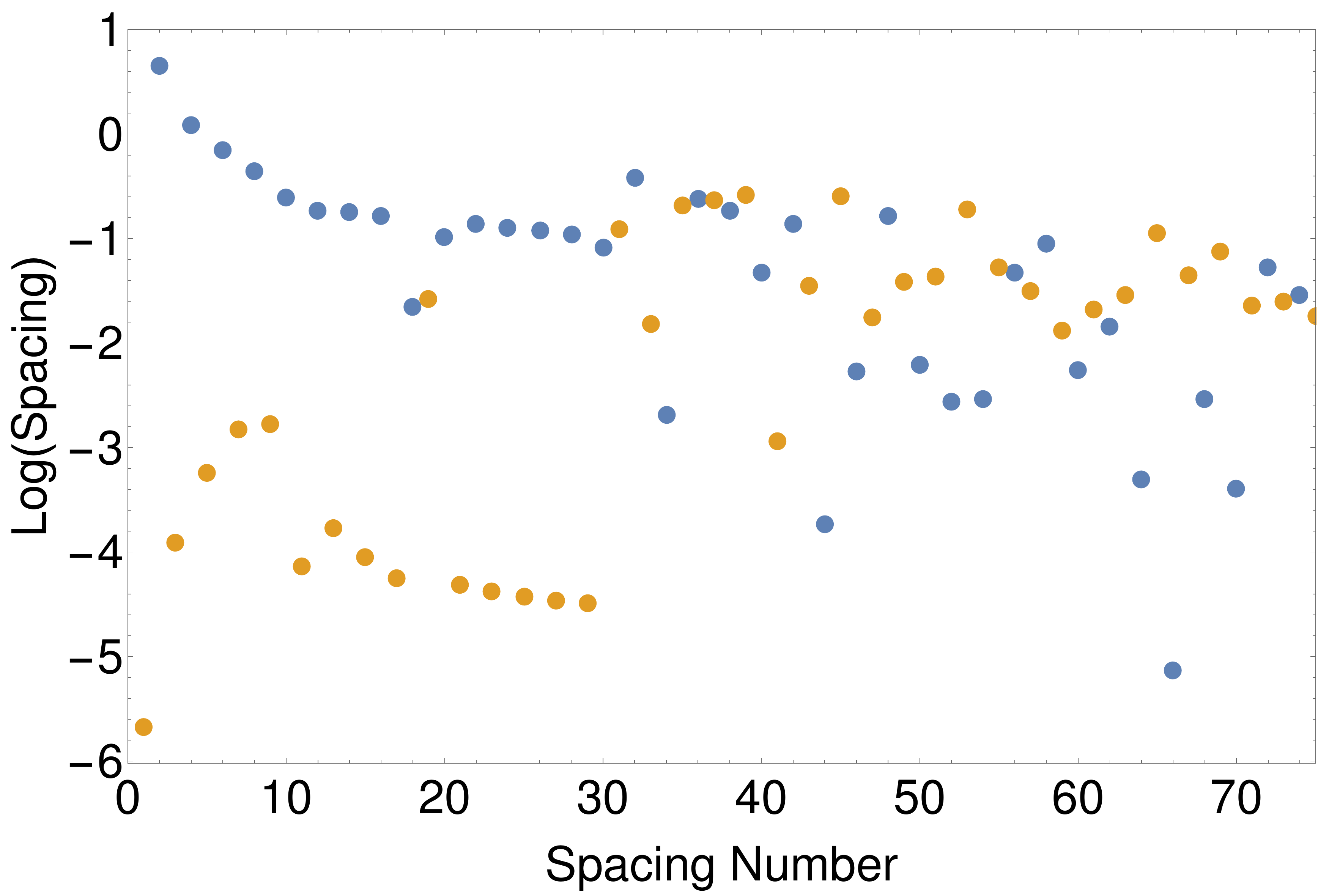}
\caption{The level spacings in the 26 site model, in the combined even and odd Ising symmetry sectors. The even spacings are blue, while the odd spacings are orange.}
\label{fig:interSpacing}
\end{figure}

Lastly, we mention an interesting feature of the lowest lying eigenstates in our model. In Figure \ref{fig:LSDOS} it is clear that there is a ``bump'' in the density of states at low energies. Based on our exact diagonalization data, at least for the system sizes we are able to consider, the size of this bump scales at best linearly with system size, and certainly not exponentially. Subject to the question of how exactly to define the edges of this bump, we find that there are 16, 18, and 19 states in this bump, in the 25, 26, and 27 site systems, respectively. For the 26 site system, we have also diagonalized the even parity, odd Ising symmetry mode, in order to study the level spacing behavior of the combined sector of energy eigenstates. Figure \ref{fig:interSpacing} shows a plot of the log of the level spacings in this combined symmetry sector, as a function of the level spacing number (there is no normalization by any mean level spacing). There is a clear alternating pattern, in which very closely spaced pairs of states are separated by energy splittings which are several orders of magnitude larger. This pattern abruptly ends outside of the bump region.

We find that the physics of these states is well described by single spin-flip product states (eigenstates of the Ising term in the Hamiltonian) which are corrected perturbatively in $g$ by the transverse-field term; the number of such states scales with system size. For these states, the energy gaps between neighboring states are large compared with $g$, and so a perturbation series in $g$ is a good approximation. For this reason, these states possess a large net magnetization. This also provides an explanation for the closely-spaced pairs of states which appear between the two Ising sectors. A perturbation series, which can be carried out strictly within one Ising sector or the other, will not result in an energy correction which differs between the two sectors until a high enough order in $g$ is reached that differences between the two Ising sectors become apparent. These differences only manifest themselves when considering product states in which the number of spin flips is on the order of half the system size (for which a combination of the Ising and spatial parity transformations could carry the product state into itself, thus excluding this state from contributing to the odd Ising sector). In order to reach such a product state from a single spin-flip state, a very high order in $g$ would need to be attained in perturbation theory, thus explaining the very small energy splitting between these states.

At energies above the scale of the single-flip product states, product states with 2, 3, 4, and higher numbers of spin flips begin to occur all at the same energy scale, so the energy splittings between states connected to each other by the transverse field term become small compared with $g$. Thus, a perturbation series fails to correctly capture the physics of the states above the energy scale of the single flip states.

We note the similarity of this alternating behavior with the behavior predicted to occur in the ``F1'' phase of similar ferromagnetic models with long-range interactions \cite{HuseSpecies}. We believe that such a thermodynamic phase should not occur in our model, since the requirements on the energy cost of a domain wall needed to produce such a phase are not satisfied. Additionally, the fact that the number of states obeying this behavior in our model does not scale exponentially with system size precludes this behavior from representing a true thermodynamic phase.

\section{Time Evolution}\label{sec:sec5}

We comment here briefly on the subject of time evolution. Given the results we have displayed, we believe that for a state prepared within the even Ising sector, below the critical energy density, the probability distribution for the order parameter should eventually time-evolve into a stationary distribution (with fluctuations about it), reflecting the thermal behavior of the broken symmetry phase: two well-separated peaks, at equal and opposite magnetization. 

There is an interesting initial product state which lives within the even Ising sector, the state in which all spins are initially polarized in the positive x-direction,
\begin{equation}
|X+\rangle = \bigotimes_{i=1}^{N} |x+\rangle_{i}= \bigotimes_{i=1}^{N} \left [ \frac{1}{\sqrt{2}}\left ( |z+\rangle_{i} + |z-\rangle_{i} \right ) \right ].
\label{eq:xplus}
\end{equation}
This initial product state has a quantum probability distribution for the order parameter given by the Binomial distribution,
\begin{equation}
P \left ( M_{z} = m \right ) = \frac{1}{2^{N}}\begin{pmatrix} N \\ \frac{m+N}{2} \end{pmatrix}.
\label{eq:xplusprob}
\end{equation}
In the large system-size limit, this distribution approaches a Gaussian centered around zero magnetization. Since we have not been able to fully diagonalize the spectrum of our 27-site model, we cannot study the exact time evolution of this initial product state in our present work. However, we note that the average energy of this initial product state is given
\begin{equation}
\langle E \rangle = -gN = -1.5N,
\label{eq:xplusen}
\end{equation}
corresponding to an energy density which is \textit{below} the critical energy density. The energy variance of this state can also be shown to scale like $\sqrt{N}$. Assuming that the results we have found here hold for larger system sizes, this suggests that the $|X+\rangle$ state is an example of an initially uncorrelated product state, with a probability distribution characteristic of the symmetric phase, which will dynamically time evolve into a state with broken symmetry, in which the order parameter probability distribution settles into a stationary two-peak structure. We believe this represents a non-trivial prediction for the time evolution of quantum Ising systems which could be tested experimentally in the near future by with an array of superconducting qubits, or in a system of trapped ions.

For states not prepared strictly within a sector of definite Ising symmetry, our results here are insufficient to predict the long-time behavior of the probability distribution of the order parameter. We expect that large, off-diagonal matrix elements of the order parameter between the two Ising sectors will generically result in long-time oscillations of the probability distribution, and that it will never thermalize in the sense of settling into a particular steady configuration. We hope to to return to this issue in future work.

\section{Summary}\label{sec:sec6}

We have studied the quantum transverse-field Ising model in one dimension with long-range interactions with a power-law decay (with an exponent of 1.5), a model that has a broken-symmetry phase at low temperature. We have seen clear signatures of eigenstate thermalization and quantum chaos within this broken-symmetry phase. We believe this represents the first time that such behavior has been seen in a clean system, without disorder, and without the need to average over multiple symmetry sectors or disorder realizations. Furthermore, under the assumption that this behavior holds for larger system sizes, we believe that this allows us to make a non-trivial prediction about the time evolution of such a system when prepared in certain initial states. We believe that this time evolution should allow for the possibility that an isolated quantum system which is prepared in an uncorrelated state can dynamically settle into a long-time order parameter probability distribution which is reflective of the broken symmetry phase. 

\begin{acknowledgments}

We thank Rubem Mondaini, Marcos Rigol, and Syrian Truong for helpful discussions, and Anders Sandvik for guidance on the SSE method.
This work was supported in part by NSF Grant PHY13-16748. We acknowledge support from the Center for Scientific Computing from the CNSI, MRL: an NSF MRSEC (DMR-1121053) and NSF CNS-0960316.

\end{acknowledgments}

\bibliography{1DLRTFIM}

\end{document}